**BP10231**

# Microscopic Reversibility, Space-Filling, and Internal Stress in Strong Glasses

J. C. Phillips

Dept. of Physics and Astronomy, Rutgers University, Piscataway, N. J., 08854-8019

ABSTRACT

The axiomatic theory of ideally glassy networks, which has proved effective in describing phase diagrams and many properties of chalcogenide, oxide, and even molecular glasses, is here broadened to describe both geometrical properties, such as the first sharp diffraction peak, and kinetic properties, such as the nonreversible enthalpy of the glass transition, as measured by modulated differential scanning calorimetry. The discussion shows why the latter is such an effective tool for identifying strong glasses.

## 1. Introduction

Almost all liquids crystallize rapidly at their melting points $T_m$, while only a few avoid crystallization and can be deeply supercooled until they solidify at the glass transition temperature $T_g$. (For example, in the earth's crust glasses occur only in lava, which has been rapidly quenched from its molten state.) Some materials form "fragile" glasses only when rapidly cooled (water quenched), while others, when pure, avoid bulk crystallization, form "strong" glasses, and crystallize only when cooled very slowly. The viscosity $\eta(T)$ of the "strong" glasses is nearly exponential in temperature (nearly constant Arrhenius activation energy $E_a$), while $E_a$ increases rapidly in the fragile glasses as T is reduced towards $T_g$. The strong/fragile dichotomy provides a useful way of distinguishing "strong" network glasses (silica, window glass, …) from "fragile" molecular glasses (polyalcohols, saccharides, fused salts, …), as well as a semiquantitative scale for comparing the thermodynamic and kinetic properties of deeply supercooled molecular liquids without utilizing detailed knowledge of their molecular structures [1].

At first it may seem that this dichotomy is almost trivial: the network glasses are connected, and can avoid crystallization by chain entanglement or distributed ring sizes (covalent forces), while the molecular glasses avoid crystallization through steric hindrance, a much weaker mechanism (polarization or Van der Waals forces). The variability of $E_a$ in fragile molecular liquids also seems easily explained: networks form clusters even near $T_m$, and their viscosity is determined by the presence of defects (surface dangling bonds) [2], whose formation energy is largely independent of cluster size. The molecular liquids, on the other hand, adopt asymmetric molecular packings as T is reduced towards $T_g$, and because Van der Waals interactions are long range compared to covalent/ionic bonding interactions, $E_a$ increases rapidly in the fragile liquids with increasing cluster size as T is reduced towards $T_g$ [3].

Attractive as these explanations appear, they are far from adequate. Practicing glass scientists have long recognized that there are "sweet spots" [optimal glass forming regions] in (for example) ternary phase diagrams of network glasses, where really strong, microscopically homogeneous glasses are formed; outside these "sweet spots" even the network glasses are partially crystallized and/or microscopically inhomogeneous (brittle), at least on a nanoscale, and are more properly termed amorphous or partially glassy. The strong/fragile dichotomy is relatively uninformative, not only for network glasses, but also even for molecular glasses: it does not distinguish the ubiquitous organic solvent glycerol, with a very low $T_g$, from other polyalcohols, or optimally bioprotective trehalose (very high $T_g$!) from other saccharides [4], and it leaves the true origins of alloy glass forming regions an intriguing mystery. Decisive insights into these mysteries have appeared only recently, through the discovery of the reversibility window in network glass alloys [5]. This window is measured by modulated differential scanning calorimetry (MDSC), which separates the enthalpy of the glass transition $\Delta H(x)$ into two parts, reversible $\Delta H_r(x)$, and nonreversible, $\Delta H_{nr}(x)$, as functions of composition x.

The reversibility window apparently incorporates almost all the properties of strong glasses. Within the window $\Delta H_r(x)$ varies smoothly, much as it did outside the window, while $\Delta H_{nr}(x)$ is very small compared to its values outside the window. In fact, the compositional dependence of $\Delta H_{nr}(x)$ is qualitatively similar to that of $\Delta E_a(x,T_g)$ = $dln\eta(x)/d(1/T)$, but while the latter varies by only a factor of 2 within the combined strong/fragile glass-forming composition range of many chalcogenide glass-forming alloys, $\Delta H_{nr}(x)$ varies by a factor of 20 (see Fig. 1) [5,6]! Moreover, while the boundaries of the strong glass formers were so fuzzy in the liquid that the reversibility window was never recognized in viscosity data, those boundaries are quite abrupt in $\Delta H_{nr}(x)$, so much so that one is tempted to refer to glasses in this intermediate compositional range as the "intermediate phase" [7]. [Strictly speaking, as glasses are not in equilibrium, one cannot speak of glass phases, but as we shall see, this term is defensible in many ways.]

Characteristic features of the reversibility window include an unexpected plateau (not peak) in the molar density, confined to a composition range nearly identical to that in which $\Delta H_{nr}(x)$ is very small; anomalies in the x dependence of strong Raman band frequencies at the window's edges; a vanishing internal network stress [6] (indicative of optimized space filling, and thus related to the density plateau), and greatly reduced aging of $\Delta H_{nr}(x)$ [5]. Technologically important effects (such as photostructural transformations) are also associated with the reversibility window [8].

It seems unlikely that the commonality of these many properties in strong glasses is accidental. Could they be due to specific kinds of cluster formation in strong glasses? Here we develop simple microscopic models that explore the role of space filling (geometrical packing) in network glasses. These models were inspired by recent comprehensive studies of the first sharp diffraction peak (FSDP) in several chalcogenide alloy systems [9,10]. Structure associated with the FSDP has also been observed optically (especially by Raman spectroscopy, where it is called the Boson peak [11]). There are many qualitative macroscopic correlations between fragility and Boson peaks in diverse supercooled liquids (including polymers) [11,12], but the microscopic and quantitative origin of these correlations in network glass alloys (where the effects are 10 times larger) remains unclear. In particular, the FSDP measures some kind of cluster dimension. The data of [10] show that chemical trends in the strength and position of the FSDP are apparently independent of the reversibility window, which provides by far the most accurate and informative tests of the strong/fragile dichotomy, while the data of [9] show a close relation in a specific alloy system ($Ge_xSe_{1-x}$). Is this one case an accident, or does the FSDP have much to tell us about the nature of strong glasses?

## 2. Review of Reversibility Window, FSDP, and Chains (or Rings)

The general features of the extensive chalocogenide alloy data base are illustrated in Figs. 2-5 for two cases: (Se,Ge) and (Se,As) alloys. Figs. 1 and 2 show the reversibility window and molar densities [12,13], while Figs. 3 and 4 FSDP heights and positions [10]. Remarkably detailed studies [9] of the FSDP in $Ge_xSe_{1-x}$ alloys, based on 14 closely spaced compositions centered on the reversibility window, showed striking features (Fig. 5). Reversibility windows typically have widths of 5% and are centered on an average valence near 2.40, while the FSDP exhibits [10] wide linear variations in height and position in three well-defined regions (I-III). The reversibility window occurs in a small range near the center of region II, where there is a rich structure suggestive of a spinodal crossover [9].

FSDP's arise from a variety of mechanisms in different glasses, and are even present in non-network glasses [10]. Their temperature and pressure dependence suggest that they are decoupled from the constraining forces that determine the reversibility window [10]. In chemically ordered network glasses, including the chalcogenide alloys, the FSDP is strongest in the cation-cation partial structure factors. A simple way of visualizing the origin of the FSDP in chalcogenide alloys is that it arises as cations form rings by cross-linking chains. (High resolution Raman spectra have shown that the dominant morphology of g-Se is chains [14].) Some obvious features of the FSDP can be explained in terms of dominant building blocks [10], which are tetrahedra for (Se,Ge) and pyramids for (Se,As) alloys. These blocks initially cross link long helical Se chains (region I, Figs 4 and 5). In region II isolated building blocks form small rings, marked by alternation of corner- and edge-sharing tetrahedra in the Ge-Se case (also found in the 48-atom unit cell of high-T c-GeSe$_2$ [14]). Tetrahedral rings are presumably more rigid than pyramidal ones, so the change in slope of the peak heights is about three times larger from I to II for (Se,Ge) alloys than for (Se,As) alloys.

At the stoichiometric compositions II-III, GeSe$_2$ and As$_2$Se$_3$, there is another break in slope. In (Se,Ge) alloys the FSDP height decreases rapidly as Ge(Se$_{1/2}$)$_4$ tetrahedra are diluted by Ge-Ge contacts (especially to form ethane-like Ge$_2$(Se$_{1/2}$)$_6$), while a sharp increase in slope is seen in (Se,As) alloys as As$_4$Se$_3$ clusters are formed [7,13]. Apparently many of these chain/ring structures are retained in the supercooled liquid, and they could explain the three regions observed in the entropy of activation for viscosity of these two alloy families [15].

### 3. Microscopic Theory: Hierarchical Constraints

Constraint theory, which matches numbers of covalent bond-stretching and -bending constraints to degrees of freedom, fully explains the origin of the reversibility window, providing that the original mean-field formulation [15] is refined to include corrections associated with local stresses and the formation of self-organized isostatic backbones [16]. However, because there is so little overlap between the phenomenologies of the reversibility window and the FSDP, this theory seems irrelevant to the FSDP [10]. This and the next Section discuss the refinements of constraint theory needed to explain the FSDP, which turns out also to be a consequence of still more refined self-organization of the internal stress field.

There are several very general points that dominate the discussion. All the FSDP data [10] on five different alloys [(Se, X), X = Si, Ge, P, As, and Te] show similar trends: linear behavior between compositions where the ring structure changes. Thus the detailed local structure of a given chain segment (or rigid segment of a ring), while it affects the magnitudes of the slope changes, is incidental to the broad qualitative trends in the data. In other words, theory need not set itself the task of deriving specific ring structures by some brute force method. In fact, no such method is known at present, because it would involve solving the problem of finding optimized chain/ring structures to fill space, and even the problems of filling space with spheres, dumbbells, only corner-sharing tetrahedra, etc. are currently addressed only in numerical simulations.

The kind of medium-range order involving intact covalent bond-stretching and –bending constraints is obviously insufficient to derive space-filling by chains/rings, which involves in addition previously neglected (because weaker) long-range Van der Waals forces (dihedral angles). In a hierarchical model the neglect is apparently well justified by the many successes of constraint theory in describing most of the important physical properties of glasses. However, an N-body configuration space is quite rich, so we expect that long-range forces could produce observable effects independently.

At first sight, one is tempted to ascribe these effects to the formation of nanocrystalline clusters, that is, nanoscale phase separation. In fact, there is strong evidence [7,13] for nanoscale phase separation in many (S, X) glasses, which have distinguishably narrower reversibility windows than corresponding (Se, X) glasses, where such effects are much smaller. Moreover, the FSDP breaks between regions I and II are definitely not associated with crystallization, but merely with ring structures. Fragile glasses outside the reversibility window can begin to crystallize with small amounts of annealing[14].

Because there is so much overlap between the phenomenology of FSDP and Boson peaks, we can ask another question: What do these phenomena have in common? The answer is that both diffraction and light scattering involve *plane-wave projections*. Note that this operation need not involve actual three-dimensional crystalline periodicity: coherence between different plane waves is not necessary to observe plane-wave order or an isotropic length scale determined by a ring diameter. In other words, ring interlocking (or chain bundling) to form nanocrystalline clusters occurs at an even more refined [lower entropy] level, possibly incompatible with the residual entropies typically found in good glass formers. The average valence $N_v = 2.67$ of crystalline $GeSe_2$ lies well outside the reversibility window of (Se,Ge) alloys [$2.40 < N_v < 2.52$]; here nanocrystalline ordering can occur [14], but the reversibility window for (Se,As) alloys is [$2.29 < N_v < 2.37$], so that $As_2Se_3$, with $N_v = 2.40$, falls near the edge of the window, and it is quite easy to make glassy $As_2Se_3$ with only ring order and little crystallinity.

**4. Space Filling: Corner-and Edge-Sharing Tetrahedra**

Silica has by far the most studied and best understood FSDP. The neutron FSDP is highest when the mean Si–O–Si angle equals 146°, and the FSDP decreases by 50% if the value deviates by only 5° from the maximum. Annealing experiments [17] showed that relaxation of a quenched silica sample dramatically sharpen the FSDP. If it is assumed that relaxation alters the height of the FSDP only through a change of the mean Si–O–Si bond angle, then the largest change in the FSDP corresponds to an increase in the mean bond angle of 2.1°. The sharpening of the FSDP indicates a decrease in the width of the Si–O–Si bond angle distribution. This bond angle can be regarded as the first broken constraint in a hierarchical constraint model, so that these calculations and experiments are consistent with decoupling of secondary space-filling effects from the primary glass-forming bond-stretching and O-Si-O bond bending constraints. A curious feature of space filling with corner-sharing tetrahedra is that the range of medium-range order is much the same regardless of the corner-sharing bond angle, which is quite different in ionic $ZnCl_2$ compared to silica and $GeO_2$ [18].

Because the ring size and the Si–O–Si bond angle distribution are coupled in silica, where the tetrahedral network is based entirely on corner-sharing, one cannot say whether ring size or the width of the first broken constraint determines the FSDP. Here the data on (Se,X) glasses, X = Ge,Si, are instructive. Whereas in $SiO_2$ one finds only corner-sharing tetrahedra, in c-XSe$_2$ the tetrahedral contacts are entirely edge- sharing (X = Si), or half edge- and half corner-sharing (X = Ge). However, the alloy trends in the FSDP are virtually identical in the two cases [10], showing that there is an underlying factor (which must be related to space filling by tetrahedra) that is more basic than local statistics. Although the reversibility windows are also virtually the same for (Se,X) glasses, X = Ge,Si, because the range spanned, $0.20 \leq x \leq 0.26$, covers only half of region II and only 1/5 of the entire network-forming regions I-III, it is clear [10] that the isostatic stress condition responsible for reversibility windows [6] does not determine the ranges of the FSDP. However, as already recognized in [9], the isostatic reversibility window can coincide with the quasi-spinodal region of the FSDP.

Many early molecular dynamics simulations did not obtain an FSDP in agreement with partial structure factors (only cation-cation contributions to the FSDP), but several more recent simulations have been successful for (Se,X) glasses, X = Ge,Si. These studies found that conventional cook and quench preparations yielded too "liquid-like" structures, but that starting from various *different*, but specifically tetrahedral space-filling, topologies yielded qualitatively similar and qualitatively satisfactory results [15, 19,20]. The most detailed study, for g-SiSe$_2$, is based on a hybrid model that started from two configurations, one liquid and one the edge-sharing crystalline structure [15,20], that deliberately introduces a strong "memory" of the latter, which persisted in the final results. These showed that the edge-sharing units made the largest contribution to the FSDP.

More quantitative models can be obtained by careful study of Raman and magic-angle spinning NMR data [21-23]. The most informative data are the detailed and extensive Raman spectra for Se$_{1-x}$Si$_x$ glasses [21]. The nearly symmetric A$_1$ breathing modes of corner- and edge-sharing tetrahedra are clearly separated. Their integrated scattering strengths for x = 1/3 are in a 1:3 ratio, which is in good agreement with NMR estimates [23]. Moreover, between x = 0.31 and 1/3, *both* bands develop a doublet structure, which is clearly consistent with tetrahedral chain segments containing 3 edge-sharing units [22] and 3 corner-sharing units, in other words, a two-cluster model. The NMR data suggest that the (edge-sharing)/ (corner-sharing) cluster ratio is roughly 3:1 for x = 1/3. On the other hand, for medium x ~ 0.20- 0.25, the ratio is close to 1. Thus the increase in this ratio near x = 1/3 is strongly superlinear.

With these considerations one can propose a structural model for the FSDP data for $Se_{1-x}Si_x$ glasses [10]. In region I the structure is dominated by long Se chain segments, and the FSDP peak height is nearly constant [10]. Thus the FSDP is related to the interchain spacing, possibly a diameter of a transverse ring passing through several cation-cross-linked chains. As x increases, this spacing decreases rapidly as the cation cross-linking pushes the chains apart. (This occurs slightly more rapidly for X = Ge than for X= Si, as Ge is slightly larger than Si.)

A crossover occurs at the boundary between between regions I and II. In the intermediate region II the FSDP peak height increases rapidly and linearly up to x = 1/3, with no indication of the superlinear changes that occur in the Raman spectra near x = 1/3. At the same time, the slope of the decrease in the position of the FSDP, $dQ_1/dx$, decreases, and now the Si slope is slightly larger than the Ge slope. Now the FSDP can measure a "suitably averaged" longitudinal chain length. The average is taken over the two kinds of chain segments, edge-sharing and corner-sharing. These have similar lengths, and presumably form bundles. The average is dominated by the longer segments of each kind (more coherent diffraction). The dominant length changes more slowly with x for X = Si than for X = Ge because edge-sharing chain segments are more rigid than corner-sharing ones, and there are more of the former than the latter for X = Si than for X = Ge. (In general the redundancy implied by edge sharing stiffens local topologies compared to the more flexible corner sharing.) Note that in region II the presence of residual Se chain segment bundles merely dilutes the tetrahedral chain segment bundles and accounts for the overall linear increase in FSDP height with x. Of course, in region III, where the Se chain segments are replaced by units containing X-X bonds, the chemically ordered chain segment morphology is disrupted, and the height of the FSDP decreases rapidly.

While the effects of space filling can generate an FSDP in many ways (especially when the formation of an FSDP is forced, as in reverse Monte Carlo models [24]), the hybrid half liquid- half crystal model [15,20] for g-SiSe$_2$ provides a simple test for the relation $Q_1 = 2\pi/R_2$, where $R_2$ is the second neighbor cation-cation (Si-Si) spacing. The Si-Si pair correlation function for this model is shown in Fig.4 of [21], where it is separated into edge-and corner-sharing components (Si-Si)$_e$ and (Si-Si)$_c$. (Si-Si)$_e$ shows sharp first neighbor (3.0 [0.2 FWHM]A) and broader second neighbor (6.0[0.8 FWHM]A) peaks, whereas (Si-Si)$_c$ shows a broader first neighbor (3.0[0.6 FWHM]A) peak and a very broad second neighbor (6.0[1.6 FWHM]A) peak. Thus the (Si-Si)$_e$ correlations dominate the FSDP (see Fig. 5 of [21]), and from $Q_1 = 2\pi/R_2$ they predict $Q_1 = 1.05$ A$^{-1}$, in excellent agreement with experiment, Fig.5 [10]. However, note that the sharpness of the simulated model FSDP is not predicted by the sharpness of the simulated model second neighbor peak. The sharpness of the FSDP is due to planar structure built into the crystalline half of the model. The presence of such large nanocrystalline clusters for r = 2x + 2 = 2.67 in these glasses is consistent with the upper edge of the reversibility window, which occurs at x = 2.52.

**5. FSDP's and Spinodal Phase Separation**

The pronounced (cation-cation)$_e$ topology is most conspicuous in (Se,Si) glasses, but Raman spectra show that edge-sharing cation units are also present [9,15,21,24] in (Ge,Se) glasses, with the important difference that the edge-sharing units are present only in isolation, presumably cross-linking corner-sharing chain segments. Alternatively, high resolution X-ray data suggested that the FSDP in (Ge,Se) glasses originates from tetrahedra sharing a Se dimmer with a 90° bond angle [25]; this explanation is implausible, because the concentration of Se dimers must decrease as x increases, contrary to the increasing trend in the height of the FSDP.

The detailed data [9] on the FSDP for (Ge,Se) glasses are drastically different from those (24 alloy compositions) obtained by other workers [25]. These data show no flat region in $Q_1(x)$ for $0.19 \leq x \leq 0.26$, only a small plateau in the integrated area of the FSDP for $0.19 \leq x \leq 0.23$. Moreover, the data of [9] used an X ray energy (Cu Kα) that does not enhance the Ge signal, whereas the data of [25] utilized two energies near the Ge edge. Does this mean that the data of [9] are questionable?

The answer to this question appears to lie in small but crucial differences in sample preparation. The boiling point of Se is 250K lower than the melting point of Ge, so there is a strong tendency towards phase segregation in (Ge,Se) alloys. To minimize this tendency, yet still quench rapidly enough to avoid partial crystallization, the quenching temperature is usually set 50K-100K above the liquidus, but in [9], all samples were quenched from $T_q$ = 1000K, where there is substantial nanoscale phase separation [14]. These factors explain why the $T_g$'s of [9] are typically 30K-40K lower than those of other workers for $Se_{1-x}Ge_x$ alloys with the same x. At these lower $T_g$'s the configurational entropy is reduced, and the ordering of the units contributing to the FSDP can be greatly enhanced. Apparently in the samples of [9], this enhancement was just large enough to expand the small FSDP area plateau [25] to the point that the position d = $2\pi/Q_1(x)$ develops a plateau that matches the range of the reversibility window. This is an interesting result in its own right, as it shows that nanoscale phase separation is enhanced in the reversibility window, where the internal network stress vanishes [6]. It seems likely that nanoscale phase separation in (Ge,Se) alloys would led to compositional grading, with Ge enrichment in sample cores, and Se enrichment near the surface (which will melt at lower $T_g$'s). Such grading would explain why Raman spectra, taken with absorbing green light, showed a very strong Boson peak, and much weaker bond-stretching bands [9].

The variability of the spinodal width in the data of [9] and [10] raises some interesting questions. What would happen to this width if $T_q$ were 1100K or 1200K? In other words, is the good matching of the spinodal width to the width of the reversibility window shown in Fig. 5 accidental, or does it hold over a range of $T_q$? Altogether it appears that the FSDP could be used as a general method for studying surface segregation and nanoscale phase separation in a wide range of glassy alloys, especially in cases where such inhomogeneities would have desirable technological properties.

It may appear at this point that the accidental coincidence of the FSDP position plateau with the reversibility window in nanoscale graded samples is of little fundamental importance. However, the reversibility window is a universal property of strong glasses, while the FSDP is a powerful probe of the internal structural length scale of the glassy network in the Cartesian space of *plane-wave projections.* In general glassy networks do not exhibit correlations between the phase diagrams of their length scales and their physical properties, so this "accident" is actually of fundamental interest. In particular, the results of [9] demonstrate that the intermediate "strong glass" region spanned by the reversibility window has the nature of a mechanical spinodal immiscibility dome [26]. This dome is similar to the thermal dome contained in the cubic Van der Waals (V,T) equation of state for liquids and gases, with its unstable region between the two extrema. In continuum theory this dome has a characteristic (coherence) length $\Lambda(x)$ associated with the spinodal separation [26] that follows quenching from a homogenous equilibrium state into another homogenous but nonequilibrium state. The microscopic study of spinodal separation in solids first began with metallic alloys; more recently, attention has turned to applications of the Cahn-Hilliard theory [26] to colloids and polymer mixtures. Although the theory indicates that $\Lambda(x)$ is typically of order nm, direct observations of $\Lambda(x)$ are rare. Either the observed length scales are too long (colloids, polymers, and even silicate glasses [27]), or the length scales are short but the quenching kinetics too rapid to map out $\Lambda(x)$ reliably (metallic glasses [28]).

Let us return now to Fig. 5, and consider the properties of the FSDP in these $Se_{1-x}Ge_x$ alloys from the spinodal viewpoint [26]. The length associated with the FSDP, $2\pi/Q_1(x)$, is denoted by d and is shown in Fig. 5(a); apart from the plateau discussed above, it increases smoothly with x. The coherence length $\Lambda(x)$ is denoted by $D = 2\pi/\Delta Q_1$ and is shown in Fig. 5(b), together with the area of the FSDP (denoted by $\upsilon$). Between $r = 2x + 2 = 2.38$ and 2.54, $\Lambda(x)$ follows an "inverted N" pattern (the data point at 2.42 may be an error: it should be lower). The "inverted N" pattern is suggestive of a spinodal pattern with marginal mechanical stability between 2.38 and 2.42, and 2.48 and 2.54, with unstable growth between 2.42 and 2.48. In the unstable region the coherence length increases with x, while in the marginally stable regions it decreases. The decreases are caused by competition between the formation of larger isostatic units ($Q_1$ increases), which are in the minority in the marginally stable regions, and the majority units (floppy for x ≤ 2.42, stressed rigid for x ≥ 2.48). In the middle interval the isostatic units are in the majority, and Λ(x) increases as x increases. Within the limits of experimental uncertainties, the sum of the x ranges of the marginally stable regions is equal to the range of the unstable region. It appears that Fig.5, based on samples quenched from an abnormally high $T_q$ of 1000K, represents the first successful measure of Λ(x) showing these competing effects not only in the peripheral marginally stable x regions, but also in the central unstable x region. Further studies of this interesting phenomenon would be welcome.

## 6. Geometrical Packing and Strong Glasses

The most important conclusions to emerge from this detailed discussion of exceptionally rich data on network glass alloys are not themselves detailed; rather, they concern the fundamental origin of the glass-forming tendency and the nature of the strong/fragile glass dichotomy [1]. Many efforts have been made to explain glass formation entirely in geometrical terms, for instance, sphere packing. Recently the availability of powerful computers with large memories has made possible the study of the energetics of tetrahedral packing, and it has been claimed that tetrahedral packing is the basic mechanism behind the formation of strong glasses. In particular, the average potential energy in an energy landscape model has been found to contain, in addition to the usual Gaussian terms found in sphere packing, an exponential term in 1/T as T → 0, signifying the emerge of a network when the tetrahedral constraint of a coordination number N ≤ 4 is applied. This, it is claimed, describes strong glasses [29]. (Earlier it was claimed, again in an energy landscape model, that strong glasses could be explained with a Gaussian distribution of activation energies. In view of the exponentially correlated structures of deeply supercooled liquids and glasses, this claim, based on randomness, seems counter-intuitive. It is also inconsistent with many relaxation experiments, even simple dielectric relaxation [30].)

As a simple test of the validity of this model, consider the following materials: $SiO_2$, $SiS_2$, $SiSe_2$, $SiTe_2$, $GeS_2$, $GeSe_2$, $GeTe_2$, $SnO_2$. All of these satisfy the condition $N \leq 4$, except $SnO_2$, where Sn is 6-fold coordinated, but which of these materials is a strong liquid/glass? The correct answer is only $SiO_2$. The condition $N \leq 4$ correctly separates strong from fragile glasses in two cases, $SnO_2$ and $SiO_2$, but it fails in 6 cases. This success rate of 0.25 is actually much better than one finds in oxide and chalcogenide networks generally. It is as if one had a litmus paper that turned pink for acids, but it also turned pink for bases. Such a paper would measure not the $[H^+]/[(OH)^-]$ ratio, but only the presence of $H_2O$.

In fact, all strong glasses are network glasses, but as we have seen, while all network glasses exhibit FSDP, the "sweet spot" of strong glasses is a small fraction (< 10%) of the region of all network glasses. In other words, network geometry and space filling alone can explain the Cartesian FSDP, but these factors are insufficient to separate strong from fragile glasses, even for the very well studied case of network glass alloys. This domain has recently been extended from chalcogenide glass alloys to include $(Na_2O)_x(SiO_2)_{1-x}$ glasses [31], and again the "sweet spot" is found to be small, $0.18 \leq x \leq 0.23$, whereas silicate networks form over a ten times wider range $0 \leq x \leq 2/3$. (Incidentally, when lime is added to form the soda-lime-silica ternary, this sweet spot expands to include window glass [32].)

The inadequacy of geometrical models is also nicely illustrated by calculations of stretched exponential relaxation of a liquid of hard symmetrical dumbbells, which exhibited unphysical oscillations (Fig. 2 of [33]) never observed in the relaxation of symmetrical polyalcohol glass-formers [4]. The reason the geometrical models fail is that they do not properly account for noncentral forces and the internal stress in network glasses [6,14]. It is difficult to see how these factors (especially the internal stress, which is nonlocal) can be described using energy landscapes. This may mean that energy landscapes, like mode-coupling, are a poor and misleading tool for studying glasses and glass kinetics [34].

## 7. Reversibility, Strings and Traps

There have been many discussions of the origin of microscopic irreversibility, going back to Boltzman (and probably before); this subject is especially difficult in deeply supercooled liquids and glasses, where the departure from equilibrium is not first-order, but instead is exponentially large. Here we add a few remarks based on recent experimental and theoretical work, beginning with what is known about stretched exponential relaxation (SER). SER is the prototypical complex temporal property of glasses, discovered by Kohlrausch 150 years ago, and now observed almost universally in microscopically homogeneous, complex non-equilibrium materials (strong and fragile network and fragile molecular glasses, polymers and copolymers, even electronic glasses). The Scher-Lax trap model (1973) is paradigmatic for electronic SER; for molecular SER Phillips identified [34] two "magic" shape fractions $\beta = 3/5$ and $3/7$, as confirmed by many later experiments recently reviewed [35]. The predictions of the trap model for glassy SER are accurate for dozens of examples to a few %, and involve *no adjustable parameters*. This success achieved in spite of the enormous complexity of glassy relaxation, and the very wide variety of topologies in these materials, is unprecedented, and it leaves no doubt that traps are fundamental to glassy kinetics. Recently traps have become apparent in molecular dynamics simulations (MDS) of hard spheres [36] and a network glass [37]. So far, MDS have studied the structure of these traps mainly for hard sphere (very fragile) glasses [38], or in network liquids far above their melting points [39], but with the Ag-doped network glass [37] much more informative studies are possible.

In any case, because of the presence of traps, in MDS the motion of excited particles goes from trap to trap along fairly short paths that resemble strings, which has led many to suggest that glass kinetics is dominated by string-like motion [39,40]. However, all these models are qualitative, and it is apparent that the fundamental relaxation process that determines SER in glasses is still *diffusion* to traps in configuration space (when non-central forces are present, rotations are important, and configuration space is not the same as Cartesian space), as described by the original Scher-Lax model, as latter modified to include effective and ineffective relaxation channels [34,35]. In other words, it is SER and the stretching fraction β that are observed experimentally, in all microscopically homogenous glasses, and only the trap model successfully predicts these *observables*.

There are many aspects to the reversibility window [5-8,16]; all are surprising, but some are more easily explained than others. In (Se,X) glasses, X = Ge and Si, the window extends from the mean-field value of coordination number r = 2.40 to 2.52. The lower end of this range is exactly what one would expect from a model of percolating isostatic $X(Se_{1/2})_4$ tetrahedra [41]. Similarly, the shift to r < 2.40 in (Se,X) glasses, X = As and P, is explained [13] by the presence of X=Se double bonds in isostatic quasi-tetrahedral units $(Se_{1/2})_3X=Se$ (x = 2.28). The upper end of the reversibility window is close to 2.40 for both X = As and P. This implies that $(Se_{1/2})_3X$ pyramids, unlike $X(Se_{1/2})_4$ tetrahedra, do not fill space in a stress-free way. This is a gratifying conclusion, unproved geometrically, but fully consistent with the experience of crystal chemistry, where tetrahedra are commonplace and pyramids are rare.

The reversibility itself could come about kinetically as follows. On melting, defects form at cluster edges, and while a few of these are actually broken bonds, most of these should be broken bond angles (such as bending at Se centers). Providing that the internal stress of the network is small (nearly zero), the density defects associated with a few broken bonds will interact only weakly among themselves. Thus when dT/dt is reversed, the broken rotational constraints can be restored, and this will automatically line up the broken bonds, so that they are also restored. In this way the noncentral forces on the length scale d (5 A) can "guide" the clusters of dimension D (25 A) towards reversibility on the length scales d and D (the FSDP and its coherence length).

## 8. Conclusions: the Nature of Strong Glasses
We have examined the Cartesian nature of the FSDP, and seen why it is generally decoupled from strong glass kinetics, even though it implies a large coherence length.

We also saw in Section 6 that geometrical models are inadequate to explain why the strong glass "sweet spots" in phase diagrams are so small. So what factors, then, do define strong glasses? Operationally, strong glasses avoid both crystallization and nanoscale phase separation (the latter precedes the former, and is thus the critical factor). Nanoscale phase separation is strongly exothermic when driven by internal stress (which is long-range), so it is understandable why internal stress is small in the reversibility window [6]. Glasses are never completely homogeneous, and in practice can always be expected to separate into stiffer backbones embedded in a softer matrix. The stress is propagated by the backbones, so the latter should be isostatic (stress-free) [41].

So long as the inhomogeneities are incipient, the glass transition will be nearly reversible in MSDC, as discussed in Sec.7. There have been many discussions in the glass literature of incipient inhomogeneities in terms of equilibrium Gibbsian nucleation models [42]. While these discussions have often proved their utility for describing data on specific materials using adjustable parameters, they have so far failed to explain compositional trends, such as strong glass "sweet spots", in phase diagrams. This failure can be traced to the local nature of nucleation models, which are unable to analyze the *non-local* internal stress field, which is characteristically small (almost liquidic) in strong glasses.

I am grateful to Prof. P. Boolchand for several insightful discussions concerning data on chalcogenide alloy glasses.

*Postscript.* This paper has focused mainly on the issue of distinguishing between fragile and strong network glasses, as these are the best glass formers, yet less than 10% of them are actually strong glasses. One can also ask the question, how does one discuss fragility over a much wider range, including not only a few network glasses, but also many poor glass formers (polymers, metallic glasses). Chemical trends in fragility of this much wider class may (or may not) be successfully described by a mechanical criterion involving the ratio of shear and bulk elastic moduli [43]. However, this macroscopic (long wave length) criterion involving equilibrium properties does not predict the abrupt boundaries of the reversibility window in network glasses, which are the result of space filling at short wave lengths. Constraint theory is useful in showing the importance of space filling in the context of annealed oxide glasses [44].

One can go still further afield on the issue of reversibility by studying the jamming of glass beads [45]. There it was concluded that in the fluid state the distribution of shear forces on the beads is an equilibrium one, while the jammed state is not in equilibrium. This would appear to agree well with our conclusion that strong glasses are associated with a reversible glass transition (they are not jammed). However, in some respects the results on the glass beads disagree with the microscopic properties of supercooled liquids. Specifically, the compositional dependence of $\Delta H_{nr}(x)$ is qualitatively similar to that of $\Delta E_a(x,T_g) = d\ln\eta(x)/d(1/T$ [5,6] in the supercooled liquid (but the range of the former is 10 times larger than the range of the latter). Perhaps the non-equilibrium properties in the fluid state of glass beads just above the jamming transition are too small to be easily measured.

# Figure Captions

Fig.1. In several chalcogenide glass alloys the composition dependence of $\Delta E_a = d\ln\eta/d(1/T)$ near $T = T_g$ closely parallels [5] that of $\Delta H_{nr}$. Also in these strong glasses, the internal network stress is nearly zero, just as in supercooled liquids [6].

Fig. 2. . Molar volumes in Ge-Se and As-Se binary glasses show a broad minimum in the reversibility window.

Fig. 3. FSDP heights in binary Ge-Se and As-Se glasses [10].

Fig. 4. FSDP positions in binary Ge-Se and As-Se glasses [10].

Fig. 5. The position d and coherence length D (inversely proportional to the width) of the FSDP in Ge-Se glasses as a function of mean atomic coordination $r = 2 + 2x$ show spinodal-like effects in glasses quenched from a high temperature [9].

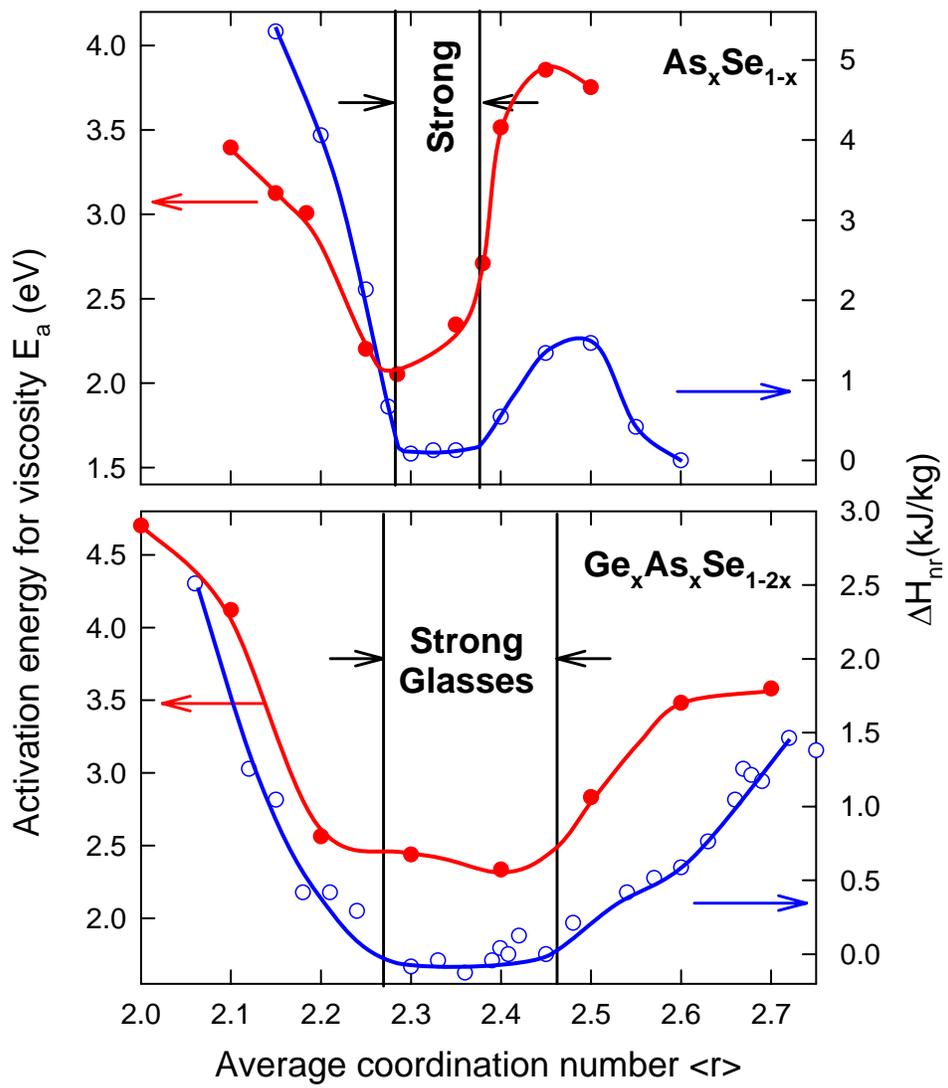

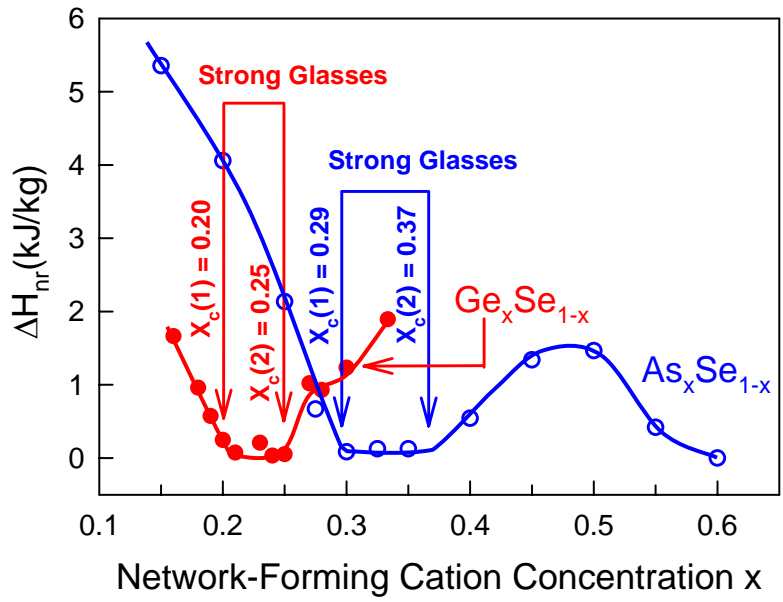

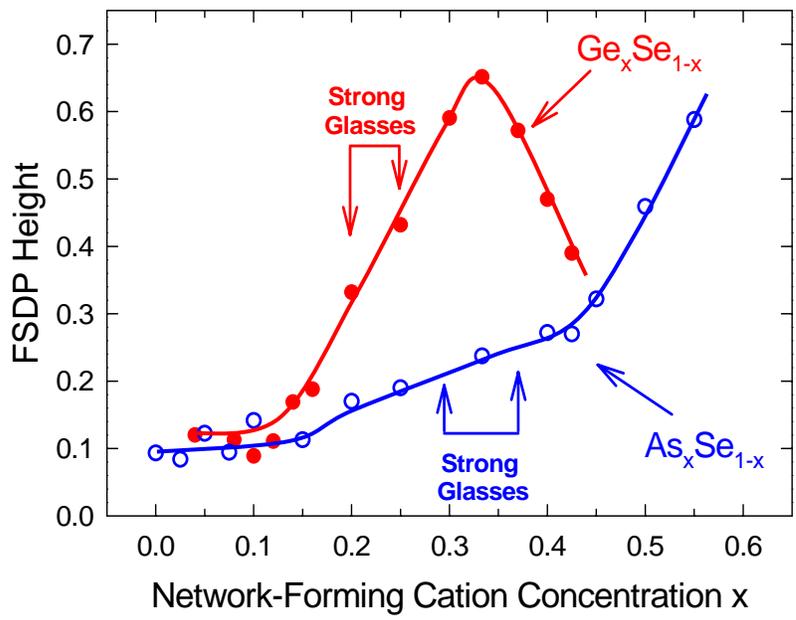

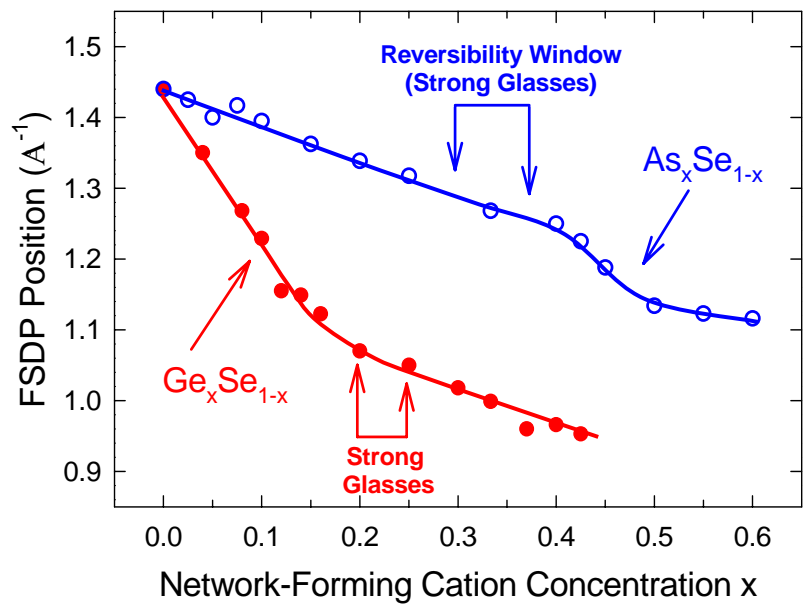

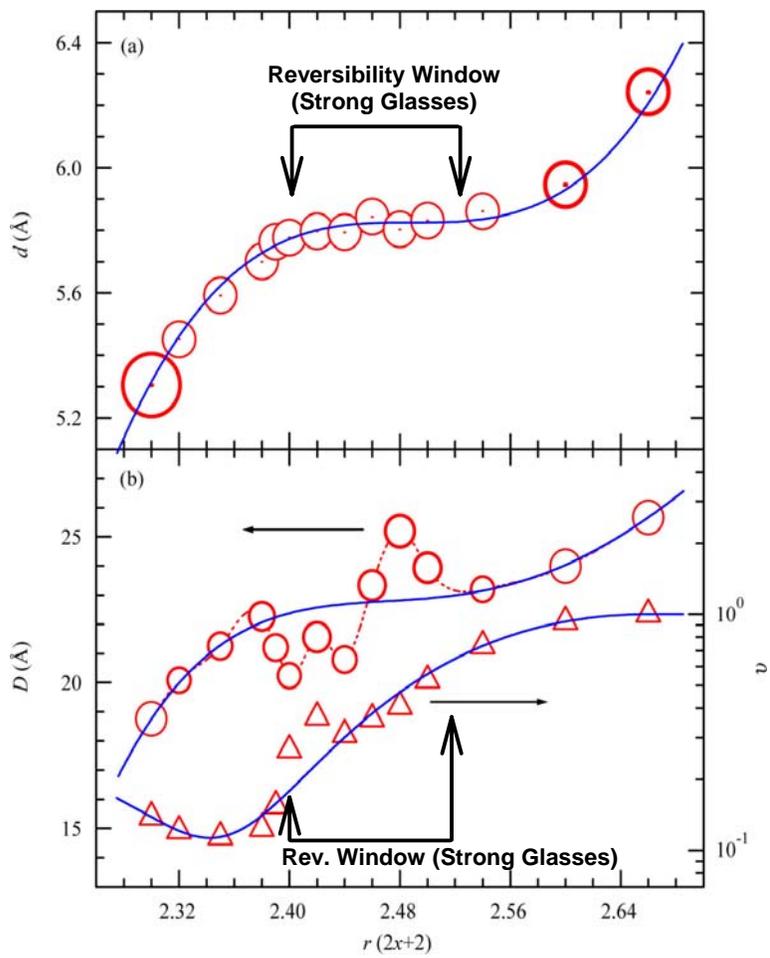